\documentclass[sigconf,nonacm]{acmart}

\usepackage{booktabs}
\usepackage{amsmath}
\usepackage{bm}
\usepackage{graphicx}
\usepackage[table]{xcolor}
\usepackage{multirow}
\usepackage{array}
\usepackage{pifont}
\usepackage{enumitem}

\setcopyright{none}
\renewcommand\footnotetextcopyrightpermission[1]{}

\newcommand{\sys}{\textsc{Parallax}}
\newcommand{\ap}[1]{\textsc{#1}}
\newcommand{\cmark}{\ding{51}}
\newcommand{\xmark}{\ding{55}}
\newcommand{\vis}{\mathrm{vis}}
\newcommand{\hid}{\mathrm{hid}}
\newcommand{\ftag}[3]{{\setlength{\fboxsep}{1pt}\colorbox[HTML]{#2}{\textcolor[HTML]{#3}{\scriptsize\textsc{#1}}}}}
\newcommand{\grp}[1]{\textcolor[HTML]{6B7280}{\scriptsize\textsc{#1}}}
\newcommand{\cc}[2]{\cellcolor[HTML]{#1}#2}
\definecolor{grprule}{HTML}{C9CED6}
\definecolor{rowhl}{HTML}{EEF3F8}

\title{Agent-Targeted Cloaking on the Web}

\author{Mu Chen}
\authornote{Corresponding author: Mu Chen.}
\affiliation{%
  \institution{Mohamed bin Zayed University of Artificial Intelligence}
  \city{Abu Dhabi}
  \country{United Arab Emirates}}
\email{mu.chen@mbzuai.ac.ae}

\author{Zhimei Chen}
\affiliation{%
  \institution{Independent Researcher}
  \country{}}

\author{Fakhri Karray}
\affiliation{%
  \institution{Mohamed bin Zayed University of Artificial Intelligence}
  \city{Abu Dhabi}
  \country{United Arab Emirates}}

\renewcommand{\shortauthors}{Mu Chen, Zhimei Chen, and Fakhri Karray}

\begin{abstract}
AI agents on the Web open and read pages for their users, analyze the content and report back. Whether this workflow is trustworthy rests on a hidden presumption: the page an agent receives at a URL is the page its user would see there. Typically, sites ask agents to present a name, such as ChatGPT-User, and decide what to show them, but a site can equally treat the \textit{agent user} differently from the \textit{real user}, denying it access or steering it to promotional or deceptive content, while the real user has no way to notice. This gives Web services a covert means to sway what users believe and choose. Publishers, CDNs and standards bodies are now writing the rules for agent traffic, yet, as our experiments reveal, existing top-leading solutions focus only on whether an agent can access a page and cannot tell whether it receives the page the real user gets. To solve this problem, we present \sys{}, a differential measurement framework that looks at each website through the eyes of an agent user and of a real user at the same moment. It compares what a site serves to a declared agent with what it serves to an otherwise identical browser, separates deliberate treatment from the natural churn of pages, and traces every difference to the signal that triggers it, keeping false alarms below one percent while recovering 97\% of planted differences. Applied to the most popular websites and a long-tail sample, \sys{} reveals that today's agents are held back mainly by their own name: about one in five sites refuses a declared AI agent that it would serve as a browser, these refusals are set by a few intermediaries and stated in no published policy, and they change what agents tell their users on most of the affected pages, while a verified fetcher passes most of them, which makes verifiable identity the foundation of a trustworthy agentic Web. \sys{} gives site operators, vendors and regulators a reproducible way to audit what the Web serves to any automated reader, and its code, full implementation and results will be released upon acceptance.
\end{abstract}



\begin{document}

\maketitle

\begin{figure}[t]
  \centering
  \includegraphics{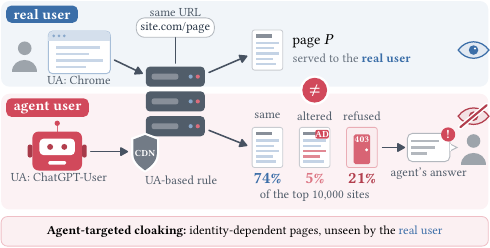}
  \caption{A site can treat the agent user differently from the real user at the same URL, and the real user sees only the answer. Against a browser that differs only in the User-Agent, ChatGPT-User gets the same page on 74\% of the top sites, an altered page on 5\% and a refusal (HTTP 403) on 21\%.}
  \Description{Two lanes. A real user's browser and an agent user fetch the same URL from a site behind a CDN with a User-Agent based rule. The real user receives page P. The agent user receives the same page on 74\% of the top 10,000 sites, an altered page with an ad on 5\%, and a 403 refusal on 21\%, and the real user only sees the agent's answer.}
  \label{fig:motivation}
\end{figure}

\section{Introduction}

Recently, there has been a trend of delegating Web reading to AI agents. Answer engines fetch pages at query time and cite them, and agentic browsers navigate sites and complete tasks on a person's behalf~\cite{venkit2025search,xu2026overviews,yang2025agenticweb}, so that the agent opens and reads a page, analyzes it and reports back to its user. This delegation is trustworthy only under a presumption that is rarely stated: the page the agent fetches is the page its user would have seen at the same URL. We call the agent the \textit{agent user} and the person it serves the \textit{real user}. Typically, a site recognizes the agent user by a declared name. Vendors' fetchers announce themselves with dedicated User-Agent tokens, publish their address ranges and honor the Robots Exclusion Protocol~\cite{anthropic2026crawl,koster2022rfc9309,openai2026crawlers}, so that each site can decide whether to admit them and what to show. Publishers, CDNs and standards bodies are now writing the rules for agent traffic, from blocking by name and charging per crawl to signed agent identity~\cite{allen2025paypercrawl,meunier2026httpsig}, and top-leading solutions measure this traffic from three angles: policy audits read what sites declare toward AI crawlers~\cite{hoffmann2026robots,longpre2024consent}, access measurements record whether a declared crawler is answered~\cite{liu2025somesite,steinacker2026misinformation}, and content scanners search crawled pages for text addressed to AI readers~\cite{brunner2026threats,khodayari2026wild}. All three observe the Web through one identity at a time, enough to ask whether an agent user is admitted but not whether it receives the page the real user gets, and reports from 2026 of sponsored copy, a defamatory biography and decoy pages served only to AI fetchers~\cite{schmalbach2026time,tatoris2025labyrinth,vlahov2025atlas} concern exactly this difference between the two users.

However, one concern naturally arises that \textit{the name that lets a site recognize the agent user also lets it treat the agent user differently from the real user, who sees only the answer and cannot notice}. We call this agent-targeted cloaking. Search-engine cloaking had a witness in the user who clicked through to the cloaked page~\cite{invernizzi2016cloak,wang2011cloak}. An agent user has none, because it is the reader and the real user receives only its answer, which gives Web services a covert means to deny the agent user access or steer it to promotional or deceptive content, and thereby to sway what the real user believes and chooses. Figure~\ref{fig:motivation} tests the presumption directly. Against a browser that differs from it only in the User-Agent string, a self-declared ChatGPT-User receives the same page on 74\% of the Tranco top 10,000 sites, another page on 5\% and a refusal on 21\%. More importantly, even the refusal is keyed on the name alone: with the engine, the address and the TLS fingerprint held fixed, the AI tokens are denied on 18--28\% of sites, a Firefox token on 0.5\% and an unknown bot on 5.7\%, and the vendor's verified fetcher, which presents the same token from a published address, receives 77\% of the pages that refuse our self-declared copy. The presumption thus fails mostly at access and only rarely in content, and it fails for the honest agent that cannot prove its name. These solutions miss both failures, and the oversight leads to three errors. A refusal enforced at the CDN but never declared in robots.txt goes unseen by a policy audit, as do 88\% of the fetcher denials we measure. The natural variation of a page between two fetches is mistaken for differential treatment, since an uncalibrated comparison flags 7.2\% of sites by chance, as many as the AI tokens. A refusal of the client or the address is blamed on the agent's name, which points the remedy at the wrong party.

In this work, we present \sys{}, a differential measurement framework for agent-targeted cloaking, and we rethink the measurement of how sites treat agents in two aspects: \textit{what difference a site makes because of the declaration}, separated from how a page varies between any two fetches, and \textit{what that difference keys on and does to an agent}. First, \sys{} fetches each page as an agent user and as a real user, emulated by an otherwise identical browser, at nearly the same moment, and uses repeated browser fetches to learn how much the page changes on its own, so that only differences beyond this natural churn count. Second, \sys{} varies the signals a site can act on, namely the declared name, the client software and the network, and compares with the vendor's verified fetcher, so that each difference is traced to its cause. Third, \sys{} sorts each difference into the form it takes, from an outright refusal to text written only for models, and tests how each form changes the answers of agents, because a refused page and a hidden instruction harm the real user in different ways.

Our work exhibits several compelling facets. First, it offers the agentic Web a reproducible way to measure what sites serve to agents, one that vendors and auditors can rerun from their own networks. Second, it overturns the expectation that agents are shown a different Web: what they meet is denial, set by a few intermediaries on the strength of a name that cannot be verified, so that declaring oneself honestly becomes a reason to be refused, while content written only for agents is rare and travels through open channels. Third, on the most popular websites and a long-tail sample, it shows that an agent user declared as AI is refused on about one in five sites, that most of these refusals appear in no published policy and vanish once the agent claims to be a browser, and that they change what agents tell their users on most of the affected pages. These findings hold across fifteen declared identities, several clients and networks, and two independent judges, and the calibrated contrast keeps false alarms below one percent while recovering 97\% of the differences we plant in a controlled testbed.

\section{Related Work}

\paragraph{Web Cloaking and Spam}
Cloaking was described as a web-spam technique two decades ago~\cite{gyongyi2005webspam,wu2005cloaking}, measured by comparing a crawler's copy of a page with a browser's copy~\cite{chellapilla2006cloaking,wu2006semantic}, tracked over time in search results for lucrative queries~\cite{wang2011cloak}, and classified into the signals that cloakers key on, from the User-Agent string to the address and to JavaScript execution~\cite{invernizzi2016cloak}. Later work moved the comparison to the client~\cite{duan2017cloakercatcher} and found cloaking logic running in the browser itself~\cite{zhang2021crawlphish}. The earliest systems already compared a crawler-versus-browser difference against a crawler-versus-crawler difference~\cite{chellapilla2006cloaking,wu2005cloaking}, and all of them presume that the second view belongs to a person who will eventually see the page. We keep the comparison, make the reader an agent, calibrate against a null on the same pages, and attribute it across signals.

\paragraph{Differential Web Measurement}
A line of measurement research asks whether the Web looks the same to every visitor: to users of anonymity networks~\cite{khattak2016anonymous,singh2017tor}, to visitors from different regions~\cite{mcdonald2018geoblocking}, to users with different profiles~\cite{hannak2013personalization}, and to crawlers with different configurations~\cite{ahmad2020apophanies,gundelach2026detecting,jueckstock2021realistic,zeber2020representativeness}, which established that a crawler's configuration changes what it records~\cite{demir2022reproducibility}. Bot-management services formalize the discrimination~\cite{azad2020webrunner,li2021goodbot,luo2026fingerprinting}. Under browser User-Agent strings, 270,000 pages showed no difference before scripts ran~\cite{intumwayase2023uaradar}, the null result against which the agent tokens of this paper are read. \sys{} turns this methodological caution into the object of study: the configuration is the agent's declared identity, and the question is how much of the Web changes with it.

\paragraph{AI Crawlers and Consent}
The spread of AI crawlers has been measured through the policies that sites publish~\cite{hoffmann2026robots,longpre2024consent}, through the compliance of the crawlers with those policies~\cite{kim2025scrapers}, and through the response a site gives to a request that carries an AI token~\cite{liu2025somesite,steinacker2026misinformation}. The last two studies record whether a page is served and find refusals that robots.txt does not declare, and a developer's probe of 300 sites under signed and unsigned identities found a handful of Markdown twins and one hidden span~\cite{zulwatha2026parity}. Proposals for a machine-facing Web, from \texttt{llms.txt}~\cite{howard2024llmstxt} and agent interfaces to sites~\cite{steiner2026mcp} to signed agent identity and content signals~\cite{allen2025contentsignals,meunier2026httpsig}, payment on fetch~\cite{allen2025paypercrawl} and Markdown negotiation~\cite{martinho2026markdown}, define how a site may address an agent, and the security literature now proposes cloaking as a defense against agents~\cite{ayzenshteyn2025cloak,li2026webcloak,liu2026expshield}. These describe policy, protocol or mechanism. We measure served content.

\paragraph{Attacks on Web Agents}
Indirect prompt injection places instructions in content that an agent will read~\cite{greshake2023indirect,liu2023promptinjection}, and benchmarks evaluate agents against such content in tool outputs~\cite{debenedetti2024agentdojo,zhan2024injecagent} and in web pages~\cite{evtimov2025wasp,liao2025eia,xu2025advagent}. Generative-engine optimization and preference manipulation show that text placed on a page can raise its visibility in an answer engine's output~\cite{aggarwal2024geo,nestaas2025adversarial,pfrommer2024ranking}, and a recent proposal describes serving such text only to agents as an attack~\cite{zychlinski2025parallel}. Two scans of the crawled Web found tens of thousands of pages carrying instructions for an AI reader, mostly crude and mostly ignored by the models tested~\cite{brunner2026threats,khodayari2026wild}, and HTML-aware detectors now target such pages~\cite{zhang2026browsesafe}. These scans see one view of each page. Our work compares the agent's view with the human's.

\begin{table}[t]
  \caption{Forms of differential treatment of a declared agent. The content forms constitute agent-targeted cloaking. Prerendering is the sanctioned variant of augmentation, and only denial is seen by the agent.}
  \label{tab:taxonomy}
  \footnotesize
  \setlength{\tabcolsep}{3.2pt}
  \begin{tabular}{@{}>{\raggedright\arraybackslash}p{66pt}>{\raggedright\arraybackslash}p{64pt}>{\raggedright\arraybackslash}p{73pt}c@{}}
    \toprule
    Form & Agent receives & Evidence & Seen \\
    \midrule
    \ftag{Den}{E5E8EB}{475363} Denial & no usable page & 4xx/5xx, timeout, challenge page & \cmark \\
    \ftag{Red}{E0EDF3}{2F6A87} Reduction & a subset of the human page & main text shorter, body removed & \xmark \\
    \ftag{Sub}{F9F0DF}{A37B2C} Substitution & a different page & main text, items, prices or locale replaced & \xmark \\
    \ftag{Aug}{E8EFE8}{527554} Augmentation & the human page plus more & extra visible or hidden text & \xmark \\
    \ftag{Pre}{ECEAF4}{675E8A} Prerendering & a server-rendered page & the browser's static copy is a script shell & \xmark \\
    \ftag{Agt}{F6E5EA}{97475B} Agent-directed & text addressed to the model & instructions, agent pages, sponsored copy & \xmark \\
    \bottomrule
  \end{tabular}
\end{table}

\section{\sys{}}
\label{sec:method}

\begin{figure*}[t]
  \centering
  \includegraphics{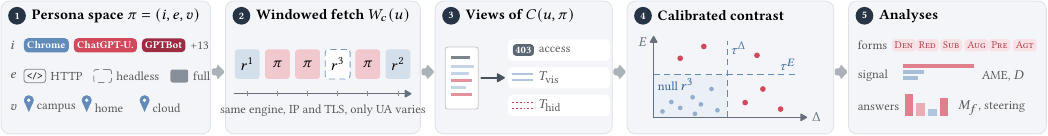}
  \caption{\sys{} fetches each page in every engine-vantage cell between Chrome reference fetches, splits responses into access, visible and hidden views, subtracts the reference variation, and passes confirmed excess to three analyses.}
  \Description{Five panels: the persona space of identities, engines and vantages; the windowed fetch that places personas between Chrome fetches r1 and r2 with a random-slot r3; the access, visible and hidden views of a response; the calibrated contrast with null thresholds; and the analyses of forms, signals and answers.}
  \label{fig:pipeline}
\end{figure*}

\sys{} answers four questions about a site: whether it treats a declared agent differently from a person (Section~\ref{sec:contrast}), what form the treatment takes (Section~\ref{sec:taxonomy}), which signal it keys on (Section~\ref{sec:attribution}) and whether the treatment changes what an agent tells its user (Section~\ref{sec:consequences}). Figure~\ref{fig:pipeline} shows the pipeline. Section~\ref{sec:prelim} formalizes the problem and Section~\ref{sec:personas} the personas and the fetch protocol.

\subsection{Preliminary}
\label{sec:prelim}

\paragraph{Problem Statement}
A site distinguishes its visitors by four classes of signal: (1) the declared User-Agent string and headers, (2) the network address and its reputation, including the vendors' published ranges, through which sites verify a crawler's claim~\cite{google2026crawlers,li2021goodbot}, (3) the transport and rendering fingerprint, from TLS settings to JavaScript execution~\cite{ahmad2020apophanies,jonker2019fingerprint}, and (4) behavior over time. Commercial bot management evaluates all four~\cite{azad2020webrunner,gundelach2026detecting}. We therefore model a visit as a persona $\pi=(i,e,v)\in\mathcal{I}\times\mathcal{E}\times\mathcal{V}$, a declared identity $i$ fetched with rendering engine $e$ from network vantage $v$, and call $c=(e,v)\in\mathcal{C}$ its cell. For a site $s\in\mathcal{S}$ with sampled pages $\mathcal{U}_s$, $C(u,\pi)$ is the response of page $u$ to $\pi$: a status, a visible text $T_{\vis}$ and a hidden text $T_{\hid}$. With $i_0$ the Chrome token, agent-targeted differential treatment of form $f\in\mathcal{F}$ (Table~\ref{tab:taxonomy}) is an indicator $y_f(u,i,c)$ that the response to $(i,c)$ differs in form $f$ from the response to $(i_0,c)$ by more than the page's own variation between fetches, and agent-targeted cloaking is its restriction to the content forms $\mathcal{F}\setminus\{\text{\ap{Den}}\}$. The quantity of interest is the share of sites that treat identity $i$ differently from the reference in cell $c$,
\begin{equation}
  \rho_f(i,c)=\mathop{\mathbb{E}}_{s\in\mathcal{S}_{i,c}}\Big[\,\max_{u\in\mathcal{U}_s} y_f(u,i,c)\Big],
  \label{eq:rho}
\end{equation}
where $\mathcal{S}_{i,c}$ holds the sites with a valid window in $c$ (Section~\ref{sec:contrast}), together with the signal of $\pi$ that switches $y_f$ on.

\paragraph{Forms and Threat Model}
Denial (\ap{Den}) withholds the page and is visible to the agent. The content forms are not. Reduction (\ap{Red}) and substitution (\ap{Sub}) exist in classic cloaking and geo-blocking~\cite{khattak2016anonymous,mcdonald2018geoblocking}. Prerendering (\ap{Pre}) is the sanctioned variant of augmentation (\ap{Aug}). Agent-directed text (\ap{Agt}) is specific to a reader that interprets text as instruction. A site, or a CDN acting for it, may vary its response on any signal, for defensive reasons, from decoy text for LLM scrapers~\cite{ayzenshteyn2025cloak,li2026webcloak,liu2026expshield} to tarpits~\cite{tatoris2025labyrinth}, commercial ones such as sponsored insertions~\cite{guaglione2026time}, adversarial ones that redirect the agent's task~\cite{greshake2023indirect,liao2025eia,zychlinski2025parallel}, or infrastructural ones such as payment and Markdown negotiation~\cite{allen2025paypercrawl,martinho2026markdown}. We assume no malice. A measurement can set any $i$, $e$ and $v$, but cannot originate from a vendor's address range or present its signature~\cite{meunier2025forgetips,meunier2026httpsig}. We therefore also ask the vendor's deployed fetcher to retrieve pages (Section~\ref{sec:personas}).

\paragraph{Shortcomings of Byte Comparison}
Fetching a page as a person and as an agent and comparing the bytes fails for three reasons. i) Pages differ between any two fetches through timestamps, rotating advertisements, A/B tests and personalization~\cite{jueckstock2021realistic,zeber2020representativeness}, so a byte comparison flags most of the Web. ii) An off-the-shelf agent persona differs from a browser in its User-Agent string, in whether it executes JavaScript and in where it connects from. A difference measured on the bundle cannot be attributed, yet the policy question depends on the attribution, because a site that keys on the User-Agent string penalizes honesty whereas one that keys on JavaScript penalizes a capability. iii) A 403 and a hidden paragraph addressed to the assistant both differ from the human page, but only the second needs an agent to be harmful. \sys{} addresses i) with a calibrated within-cell contrast, ii) with a factorial design, and iii) with a visible-hidden decomposition and a typed taxonomy.

\subsection{Persona Space and Fetch Protocol}
\label{sec:personas}

\paragraph{Factorial Personas}
The identity set $\mathcal{I}$ has 16 members (Table~\ref{tab:prevalence}): the Chrome token $i_0$ as the reference, three on-demand agent fetchers, two AI search crawlers (OAI-SearchBot and PerplexityBot), three training crawlers, two search crawlers as a control for classic cloaking, four controls that separate the treatment of agents from that of browsers and bots in general (Firefox, the \texttt{HeadlessChrome} token, a plain HTTP library token and an unknown research-bot token), and a protocol persona, a Chrome client that asks for \texttt{text/markdown}~\cite{martinho2026markdown}. The engine set $\mathcal{E}$ holds an HTTP client that, like the major vendors' crawlers~\cite{zecchini2024aicrawler}, does not execute JavaScript and replays Chrome's TLS and HTTP/2 fingerprint, Chromium in headless-shell mode, and full Chromium under a virtual display with its automation flag removed. Within an engine every identity sends Chrome's navigation headers without client hints, so that two personas of a cell differ only in the User-Agent string. The vantage set $\mathcal{V}$ holds a campus network, a residential line and a cloud datacenter in Australia, in distinct autonomous systems and in none of the eleven address-range files that the vendors publish. All 15 identities $i\neq i_0$ are swept under the HTTP engine from the campus on every page. A stratified subsample of 2,000 sites, 200 per rank decile, is also swept from the residential line and the datacenter (homepages) and, for $i_0$ and four identities (ChatGPT-User, Perplexity-User, GPTBot and Googlebot), under both rendered engines from the campus, on all 2,000 sites.

\paragraph{Windowed Fetching}
For a page $u$ and a cell $c$, the window
\begin{equation}
  W_c(u)=\big\langle\, r^1,\ \pi_{\sigma(1)},\ \dots,\ \pi_{\sigma(n)},\ r^2\,\big\rangle
  \label{eq:window}
\end{equation}
fetches the $n$ personas of $c$ in a random order $\sigma$ between two Chrome fetches $r^1$ and $r^2$, and inserts a third Chrome fetch $r^3$ at a uniformly random slot, with two seconds between requests to a host, so that a window spans one to two minutes. Pages are the homepage and up to two internal same-origin pages sampled from its links, excluding login, cart, search and legal endpoints. Each site also contributes its \texttt{robots.txt} and \texttt{llms.txt}. To test what a spoofed declaration misses, we fetch 326 URLs, stratified by our labels, through a deployed Claude fetcher whose requests carry the Claude-User token from an address inside Anthropic's published range, and record whether it receives the page that our Claude-User persona received.

\paragraph{Views of a Response}
The access view records the status class, the final URL after redirects, and signatures of challenge and block pages from the major bot-management vendors and of proof-of-work interstitials. The visible view $T_{\vis}$ is the main text of the rendered document, extracted after computing styles so that elements with \texttt{display:none}, \texttt{visibility:hidden}, zero opacity, zero font size or off-screen placement are excluded. For the HTTP engine the same extraction runs on the static DOM. The hidden view $T_{\hid}$ collects the text that the visible view excluded, together with HTML comments, \texttt{alt}, \texttt{title} and \texttt{meta} content, and zero-width or tag-block Unicode characters, which prior work identifies as carriers of text meant for a model~\cite{greshake2023indirect,rehberger2024ascii}.

\subsection{Within-Cell Contrast Calibration}
\label{sec:contrast}

\paragraph{Displacement and Excess}
A difference is attributed to the declaration only when it is measured against a reference that differs from the persona in nothing but the declaration, which within a window is $i_0$ itself. Text views are compared with a dissimilarity $\delta\in[0,1]$, one minus the Jaccard similarity of 5-gram shingle sets over normalized main text~\cite{broder1997resemblance,wu2006semantic}. Markdown and HTML responses are both reduced to main text first~\cite{barbaresi2021trafilatura}, so that a change of format alone is not a difference. The displacement of persona $\pi$ is measured against the natural variation $\nu(u)$ of the page,
\begin{equation}
  \begin{aligned}
    \Delta(u,\pi)&=\min_{k\in\{1,2\}}\delta\big(C(u,r^k),C(u,\pi)\big)-\nu(u),\\
    \nu(u)&=\delta\big(C(u,r^1),C(u,r^2)\big),
  \end{aligned}
  \label{eq:displacement}
\end{equation}
where the minimum credits the persona with whichever reference fetch it resembles more, so that drift between $r^1$ and $r^2$ does not inflate $\Delta$. Displacement measures whether the main text changed. The excess $E(u,\pi)$ measures whether text was added. It counts the tokens in sentences of the persona's visible or hidden view that occur in neither $r^1$ nor $r^2$, minus the churn of the page, the mean of the tokens that $r^2$ adds to $r^1$ and that $r^1$ adds to $r^2$, so that rotating headlines and listings cancel.

\paragraph{Null Calibration}
Because $r^3$ occupies a random slot, its displacement and excess are exchangeable with those of any persona that the site treats like the reference, and they serve as the null: if a site treats $\pi$ as it treats $i_0$, then $(\Delta,E)(u,\pi)$ and $(\Delta,E)(u,r^3)$ have the same distribution, because $r^3$ is placed uniformly among the slots of $W_c(u)$. A persona is flagged as altered when either statistic exceeds a threshold set per cell $c$ and per bin $b(u)$ of $\nu$, so that volatile pages receive larger thresholds without masking stable ones,
\begin{equation}
  \begin{aligned}
    a(u,\pi)&=\bm{1}\big[\,\Delta(u,\pi)>\tau^{\Delta}_{c,b}\ \vee\ E(u,\pi)>\tau^{E}_{c,b}\,\big],\\
    \alpha&\geq\mathop{\mathbb{E}}_{s\in\mathcal{S}}\Big[\,\max_{u\in\mathcal{U}_s}a(u,r^3)\Big],
  \end{aligned}
  \label{eq:threshold}
\end{equation}
where $b=b(u)$ and the second line bounds the share of sites that the null flags on either statistic at $\alpha=1\%$, and by exchangeability the same bound holds for every persona that the site treats like a browser. The thresholds are fit on one half of the sites and applied to the other, so that the null rate we report is out of sample.

\paragraph{Denial and Units}
A persona is denied, $y_{\text{\ap{Den}}}=1$, when it receives an error status, no response within 25 seconds, a challenge or block page of a bot-management vendor, or a short page that announces a block, while $r^1$, $r^2$ and $r^3$ all succeed (for the null $r^3$, while $r^1$ and $r^2$ succeed), after retrying 429, 5xx and network errors once. A window is valid only when all three reference fetches succeed, so sites that refuse Chrome itself from a vantage fall out of $\mathcal{S}_{i,c}$, which can only lower the denial rates we report. A site is a registrable domain, and Equation~\ref{eq:rho} counts it once when any of its pages is flagged. Every flagged page, including those the null flagged, is fetched again days later to measure recurrence.

\subsection{Divergence Taxonomy and Labeling}
\label{sec:taxonomy}

A flagged page receives one content form, the first that applies in the order of severity \ap{Agt}, \ap{Red}, \ap{Sub}, \ap{Pre}, \ap{Aug}. A rule-based detector assigns \ap{Red} when the persona's main text is shorter and contained in the reference, \ap{Aug} when the reference is contained in a longer persona text or when only the excess is flagged, \ap{Pre} when the reference's static page is a script shell and the persona's is not, and \ap{Sub} otherwise. Form \ap{Agt} is excess that matches a lexicon of constructions drawn from the prompt-injection and generative-engine-optimization literature~\cite{aggarwal2024geo,greshake2023indirect,khodayari2026wild,nestaas2025adversarial}: address to an assistant or agent, imperatives about instructions or prior context, preference statements relative to alternatives, fake system framing, and requests that would carry the agent's context off the page. A negotiated response whose main text matches the reference is excluded. The detector is measured in two ways. A planted testbed of 1,480 pages built on 40 real article texts applies each content form at three magnitudes (a few words, a paragraph, most of the page) and, for added text, in five channels (visible, CSS-hidden, comment, attribute, Unicode tag characters) only to AI tokens, and runs through the same harness and thresholds, which gives recall per form and channel and the false-flag rate on unmodified pages and non-AI identities. In the wild, a language model reading the excess as delimited untrusted data~\cite{hines2024spotlighting} triages every agent-only span into benign variation, browser notices, rotating items, text written for AI systems, sponsored content and instructions to AI, and every span outside the first three classes is reviewed by hand.

\subsection{Signal Attribution}
\label{sec:attribution}

The within-cell contrast isolates the declaration. Attribution asks how the declaration compares with the signals an agent cannot change. On the subsample, every persona of every cell is compared with the human reference $h=(i_0,\text{headful},\text{campus})$, which yields a denial indicator $y^h(s,\pi)$ per site and persona, and we fit a mixed-effects logistic model,
\begin{equation}
  \operatorname{logit}\Pr\big[\,y^h(s,\pi)=1\,\big]=\xi_s+\zeta_{d(s)}+\bm{\beta}^{\top}\bm{x}_i+\bm{\gamma}^{\top}\bm{x}_e+\bm{\eta}^{\top}\bm{x}_v,
  \label{eq:glmm}
\end{equation}
where $\xi_s$ and $\zeta_{d(s)}$ are random intercepts for the site and for its CDN or bot-management provider $d(s)$, which enforces most denials, and $\bm{x}_i$, $\bm{x}_e$, $\bm{x}_v$ are indicators of identity, engine and vantage. We report average marginal effects in percentage points, which are comparable across factors. Alteration is not modeled across engines, because a static and a rendered copy of the same page differ in text for every identity. It is measured within cells. What an agent would regain by declaring a browser is the spoofing dividend
\begin{equation}
  D(i,c)=\Pr\big[\,y^h(s,(i_0,c))=0\ \big|\ y^h(s,(i,c))=1\,\big],
  \label{eq:dividend}
\end{equation}
the share of an identity's denials relative to $h$ that the User-Agent string alone explains, and so what an honest declaration costs it.

\subsection{Consequences for Agents}
\label{sec:consequences}

Whether a different page matters is tested on agents. For a stratified sample of flagged pages, an agent backbone $A$ answers two requests an answer engine serves, a three-sentence summary and a question about what the site offers, from three views: the browser's first fetch $r^1$, the browser's last fetch $r^2$, and the persona's view $\pi$. A judge model $J$ from a different vendor decides whether a user would come away with materially different information, and the effect of form $f$ is the excess of material changes over those that the page's own variation causes,
\begin{equation}
  M_f=\Pr\big[\,J(A_{\pi},A_{r^1})=\mathsf{mat}\,\big]-\Pr\big[\,J(A_{r^2},A_{r^1})=\mathsf{mat}\,\big],
  \label{eq:consequence}
\end{equation}
with page-level bootstrap intervals. We use six open-weight backbones from three vendors, from 7B to 70B parameters, a judge from a fourth vendor, and a second judge to measure agreement. Agent-directed text is tested on its own: for each distinct directive that sites publish in an \texttt{llms.txt} file, the judge writes two questions on which the directive matters, and a backbone answers from the homepage and the file with the directive kept or stripped, plainly and under a system prompt that marks retrieved content as untrusted~\cite{hines2024spotlighting}. The judge, blind to the condition, rates whether the answer follows the directive.

\paragraph{Scale and Cost}
A window issues $n+3$ requests to one host, eighteen in the HTTP sweep, at most one every two seconds, so a page costs about a minute of wall-clock time at its host while thousands of hosts proceed in parallel. The campus sweep of 19,678 pages takes about two hours on one 16-core node, extraction, contrasts and calibration take about two CPU-hours, and only the consequence study uses accelerators, about one A100 and three H100 GPU-hours, so that a vendor or an auditor can rerun the whole protocol from its own address range in an afternoon. Code, full implementation and results will be released upon acceptance.

\section{Results}
\label{sec:results}

The sweep covers the Tranco list of 18 October 2026. Of its 10,000 domains, 6,648 return an HTML page to Chrome from our vantages. The rest are infrastructure, API or redirect domains, refuse Chrome itself, or are filtered at the national level. Collapsing domains that share a registrable label, such as the 137 country domains of one search engine, leaves 4,705 distinct sites with a valid window from the campus. The campus sweep issued 354,204 requests over 19,678 pages. A second campus sweep and the re-fetch of every flagged page followed four to five days later.

\begin{table}[t]
  \caption{Sites (\%) that deny or alter each identity relative to the Chrome token in the same cell, HTTP engine. Shading grows with the rate. Residential and datacenter columns cover the homepages of the subsample.}
  \label{tab:prevalence}
  \footnotesize
  \setlength{\tabcolsep}{3.4pt}
  \renewcommand{\arraystretch}{0.94}
  \begin{tabular}{@{}llrrrrrr@{}}
    \toprule
     & & \multicolumn{2}{c}{Campus} & \multicolumn{2}{c}{Residential} & \multicolumn{2}{c}{Datacenter} \\
    \cmidrule(lr){3-4}\cmidrule(lr){5-6}\cmidrule(l){7-8}
    Group & Identity & \ap{Den} & \ap{Alt} & \ap{Den} & \ap{Alt} & \ap{Den} & \ap{Alt} \\
    \midrule
    \multirow{2}{*}{\grp{Controls}} & Browser null $r^3$ & \cc{FFFFFF}{0.2} & \cc{F2F7FA}{0.9} & \cc{FFFFFF}{0.1} & \cc{F3F8FA}{0.8} & \cc{FFFFFF}{0.2} & \cc{F5F9FB}{0.6} \\
     & Firefox & \cc{FDFEFE}{0.5} & \cc{ECF4F8}{1.3} & \cc{FFFFFF}{0.1} & \cc{EEF5F8}{1.2} & \cc{FFFFFF}{0.2} & \cc{ECF4F8}{1.3} \\
    \arrayrulecolor{grprule}\cmidrule(l){2-8}\arrayrulecolor{black}
    \multirow{3}{*}{\grp{AI fetcher}} & ChatGPT-User & \cc{C2C8D0}{21.2} & \cc{B8D5E3}{4.6} & \cc{D1D5DB}{16.0} & \cc{D7E7EF}{2.7} & \cc{DDE1E5}{11.9} & \cc{C2DBE7}{4.0} \\
     & Claude-User & \cc{CCD1D8}{17.7} & \cc{CFE2EC}{3.2} & \cc{D9DCE1}{13.4} & \cc{E0EDF3}{2.0} & \cc{ECEEF0}{6.6} & \cc{DCEAF2}{2.2} \\
     & Perplexity-User & \cc{C4CAD1}{20.3} & \cc{CBE0EB}{3.4} & \cc{D4D8DE}{15.2} & \cc{DEECF2}{2.2} & \cc{DFE2E6}{11.3} & \cc{D7E7EF}{2.6} \\
    \arrayrulecolor{grprule}\cmidrule(l){2-8}\arrayrulecolor{black}
    \multirow{2}{*}{\grp{AI search}} & OAI-SearchBot & \cc{C4CAD1}{20.5} & \cc{BAD6E4}{4.4} & \cc{D2D7DD}{15.5} & \cc{D3E5EE}{2.9} & \cc{DFE2E6}{11.3} & \cc{C5DDE8}{3.8} \\
     & PerplexityBot & \cc{C2C8D0}{20.9} & \cc{BCD7E5}{4.3} & \cc{D2D7DD}{15.7} & \cc{D7E7EF}{2.7} & \cc{DDE1E5}{11.4} & \cc{CBE0EB}{3.4} \\
    \arrayrulecolor{grprule}\cmidrule(l){2-8}\arrayrulecolor{black}
    \multirow{3}{*}{\grp{AI training}} & GPTBot & \cc{B4BBC5}{26.2} & \cc{BED8E6}{4.3} & \cc{C2C8D0}{21.0} & \cc{D5E6EE}{2.8} & \cc{D1D5DB}{16.4} & \cc{CFE2EC}{3.1} \\
     & ClaudeBot & \cc{AFB7C1}{27.8} & \cc{C2DBE7}{4.1} & \cc{BDC4CD}{22.7} & \cc{D7E7EF}{2.7} & \cc{D7DBE0}{13.9} & \cc{C5DDE8}{3.7} \\
     & CCBot & \cc{B7BEC8}{25.3} & \cc{C3DCE8}{3.9} & \cc{C2C8D0}{20.9} & \cc{DBE9F1}{2.3} & \cc{D1D5DB}{16.0} & \cc{CDE1EB}{3.2} \\
    \arrayrulecolor{grprule}\cmidrule(l){2-8}\arrayrulecolor{black}
    \multirow{2}{*}{\grp{Search}} & Googlebot & \cc{C1C7CF}{21.6} & \cc{AECFDF}{5.3} & \cc{D1D5DB}{16.3} & \cc{CDE1EB}{3.2} & \cc{DCDFE4}{12.1} & \cc{C3DCE8}{3.9} \\
     & Bingbot & \cc{C1C7CF}{21.7} & \cc{ACCEDF}{5.4} & \cc{D1D5DB}{16.3} & \cc{C7DEE9}{3.6} & \cc{DCDFE4}{12.5} & \cc{C3DCE8}{3.9} \\
    \arrayrulecolor{grprule}\cmidrule(l){2-8}\arrayrulecolor{black}
    \multirow{3}{*}{\grp{Other bots}} & HeadlessChrome & \cc{CACFD6}{18.6} & \cc{E0EDF3}{2.0} & \cc{D5DADF}{14.2} & \cc{EAF2F7}{1.3} & \cc{E4E7EA}{9.5} & \cc{E4EFF4}{1.7} \\
     & python-requests & \cc{BCC3CB}{23.1} & \cc{D7E7EF}{2.6} & \cc{CCD1D8}{17.6} & \cc{E4EFF4}{1.7} & \cc{D9DCE1}{13.3} & \cc{E0EDF3}{2.0} \\
     & Unknown bot & \cc{EFF1F3}{5.7} & \cc{D3E5EE}{2.9} & \cc{F4F5F6}{3.8} & \cc{E6F0F5}{1.6} & \cc{F2F3F5}{4.5} & \cc{DCEAF2}{2.2} \\
    \arrayrulecolor{grprule}\cmidrule(l){2-8}\arrayrulecolor{black}
    \grp{Protocol} & Markdown client & \cc{FDFEFE}{0.3} & \cc{A1C7DA}{6.2} & \cc{FFFFFF}{0.1} & \cc{B0D0E0}{5.1} & \cc{FFFFFF}{0.3} & \cc{C2DBE7}{4.0} \\
    \bottomrule
  \end{tabular}
\end{table}

\begin{figure}[t]
  \centering
  \includegraphics[width=\columnwidth]{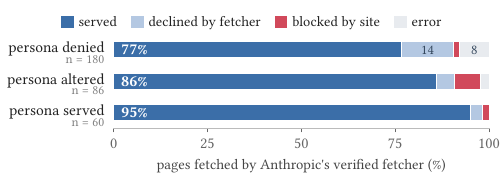}
  \caption{Anthropic's verified fetcher on pages where our self-declared Claude-User persona is denied, altered or served the browser's page. With the same token sent from a published address, the fetcher is served most of the pages that refuse the persona, and nearly every page that serves it.}
  \Description{Stacked horizontal bars for pages where the persona is denied (n = 180), altered (n = 86) or served (n = 60). The verified fetcher is served 77\%, 86\% and 95\% of these pages, with the rest declined by the fetcher, blocked by the site or errors.}
  \label{fig:verified}
\end{figure}

\subsection{Prevalence}
\label{sec:prevalence}

Table~\ref{tab:prevalence} reports, for each identity, the share of sites that deny it and the share that alter its content, each relative to the Chrome token in the same cell. The null and the browser control behave as calibrated: the random-slot Chrome fetch is denied on 0.2\% and altered on 0.9\% of sites, and the Firefox token, which changes every byte of the User-Agent string, on 0.5\% and 1.3\%. From the campus, by contrast, every declared AI identity is denied on 17.7 to 27.8\% of sites. Training crawlers are denied most. An unknown research bot, whose token names no vendor, is denied on 5.7\%, so about three quarters of the denial an AI agent meets is tied to the name it declares. Yet the spoofed search crawlers (21.6\%) and the automation tokens (18.6 and 23.1\%) are denied as often as the AI agents, so denial is not specific to AI names: it attaches to recognized automated names, of which AI agents have become one, and only at one CDN does a rule target AI names (Section~\ref{sec:denial}). Alteration is an order of magnitude rarer, 3.2 to 4.6\% of sites for the AI tokens, and most of it is shared with the unknown bot at 2.9\%: the alteration specific to AI names is 0.3 to 1.7 points. On the residential subsample every rate is lower, 13.4 to 22.7\% denial for the AI tokens, and from the Australian datacenter lower still, 6.6 to 16.4\%, with the unknown bot at 4.5\% and Firefox at 0.2\%. The ordering of identities is the same at every vantage (Spearman $\rho=0.98$). The flags are stable: when every flagged page was fetched again from the campus four to five days later, 98.6\% of the denials and 90\% of the alterations of AI identities recurred (97\% and 81\% on the residential line), against 5 of the 41 pages that the null had flagged in the first sweep. A second sweep agrees: on the 3,889 sites with a valid window in both, each identity's rates move by at most 0.3 points for denial and 0.5 for alteration, and 99\% of site-level decisions on AI identities are unchanged. The 816 sites without a valid second window deny AI tokens disproportionately (48\%), so the denominator rule lowers our rates. Beyond the top 10,000, a stratified sample of 6,000 domains ranked 10k to 1M (2,047 sites with a valid window) shows denial narrowing to training crawlers (Figure~\ref{fig:content}(b)): GPTBot, ClaudeBot and CCBot are still denied on 19.7 to 22.1\% of sites, whereas the agent fetchers fall to 7.8 to 10.9\%, the spoofed Googlebot to 10.4\% and the unknown bot to 3.5\%. At Cloudflare, which fronts half of these sites, GPTBot is denied on 25\% and ChatGPT-User on 9\%, the mark of a default rule against AI training crawlers rather than of general bot management.

\begin{figure}[t]
  \centering
  \includegraphics[width=\columnwidth]{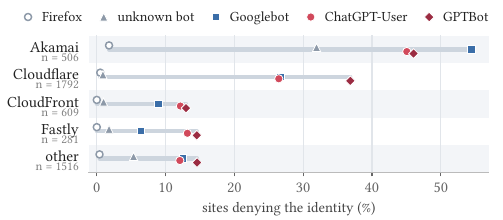}
  \caption{Sites denying each identity, by the CDN that fronts the site. An unknown bot passes Cloudflare but not Akamai. AI and search tokens are denied by both.}
  \Description{Dot plot of the share of sites denying Firefox, an unknown bot, Googlebot, ChatGPT-User and GPTBot, for sites behind Akamai (n = 506), Cloudflare (n = 1792), CloudFront (n = 609), Fastly (n = 281) and other providers (n = 1516).}
  \label{fig:cdn}
\end{figure}

\subsection{Denial Mechanisms}
\label{sec:denial}

Denial is concentrated and mechanical. 82\% of AI denials are HTTP 403 responses, 3.9\% are 402 responses that ask for payment per crawl, and 7\% are timeouts. Cloudflare and Akamai front 49\% of the sites but account for 75\% of the denials, and 78\% and 62\% of the denials behind them carry their challenge or block signature, so the rule is enforced by the intermediary even when the origin chose it. The two behave differently (Figure~\ref{fig:cdn}): sites behind Akamai deny the unknown bot on 32\% and ChatGPT-User on 45\%, the mark of generic bot management, whereas sites behind Cloudflare deny the unknown bot on 0.9\% and ChatGPT-User on 27\%, the mark of a rule written for AI names. Denials also move together: 1,522 sites (32\%) deny at least one AI identity, 726 deny all eight, 350 only training crawlers and 16 only on-demand fetchers, so about half of the denying sites set their stance once for every AI name. By category (Table~\ref{tab:categories}), news and adult sites deny training crawlers more often than the spoofed Googlebot (28 and 40\% against 18\%), retail, travel, finance and government sites refuse every automated name, and developer documentation denies least.

The verified fetcher shows whom sites refuse (Figure~\ref{fig:verified}). We asked Anthropic's deployed fetcher, through the product's fetch tool, to retrieve 326 URLs, 180 of them pages that deny our self-declared Claude-User, and coded each result as content, a block page, a refusal by the fetcher (robots.txt or opt-out), or an error. On a server of our own, the fetcher sent the Claude-User token from Anthropic's published range, so it differs from our persona in address and client stack. It received 77\% of the denied pages (77\% after reweighting to the CDN mix of all denials), 90\% of those behind Akamai and 80\% of those behind Cloudflare. Another 14\% were declined by the fetcher itself, and only 2\% were blocked by the site. On the 60 pages that serve our persona it received 95\%. On most pages, therefore, what is refused is the unverified claim to be the agent, the treatment that a request naming Googlebot from an unlisted address has long received~\cite{google2026crawlers}, and our denial rates are an upper bound on what this vendor's verified fetcher meets.

Declared policy and enforced treatment diverge in both directions. Robots.txt disallows the on-demand fetchers on 7\% of sites and the training crawlers on 13\%. 62\% of the sites that disallow a fetcher still serve it, leaving the decision to the fetcher, whose vendors differ in whether on-demand fetches follow robots.txt~\cite{anthropic2026crawl,openai2026crawlers}. In addition, 88\% of the denials of fetchers come from sites whose robots.txt does not disallow them, the silent refusal that Liu et al.~\cite{liu2025somesite} observed for Anthropic's crawlers, now across every AI identity. The rule that denies is thus a rule about impersonation enforced at the edge rather than the site's stated policy.

\subsection{Content-Level Differences}
\label{sec:content}

\begin{figure*}[t]
  \centering
  \includegraphics{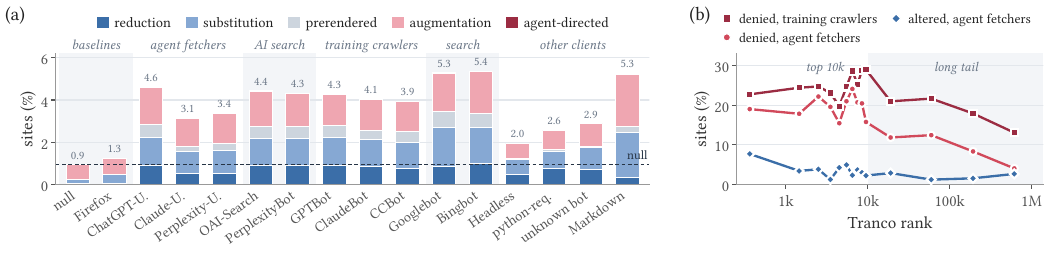}
  \caption{Content-level differences. (a) Sites (\%) showing each form toward each identity, each site counted under its most severe form. (b) Denial and alteration by Tranco rank, by decile in the top 10,000 and on a sample of ranks 10k to 1M (shaded).}
  \Description{Panel (a): stacked bars of reduction, substitution, prerendered, augmentation and agent-directed forms for each identity, between 0.9\% for the null and 5.4\% for Bingbot. Panel (b): lines of denial of training crawlers, denial of agent fetchers and alteration of agent fetchers across Tranco rank from 1k to 1M.}
  \label{fig:content}
\end{figure*}

Figure~\ref{fig:content}(a) assigns each site and identity its most severe content form. For every AI identity the content forms together reach 3.2 to 4.6\% of sites, against 2.9\% for the unknown bot. Reductions and substitutions, typically a shorter or different rendering for any non-browser string, occur about equally for the AI tokens and the unknown bot. The excess lies in two forms. Prerendering, in which the browser's static page is a script shell and the crawler receives a server-rendered page, reaches 0.2 to 0.6\% of sites for the AI tokens and 0.1\% for the unknown bot, the signature of dynamic-rendering services keyed on crawler names. Augmentation, in which the agent receives text that neither browser fetch contains, exceeds the unknown bot by up to 0.7 points. Alteration is highest among the top 1,000 sites (Figure~\ref{fig:content}(b)), where large platforms keep separate renderings for bots. On the planted testbed (Figure~\ref{fig:recall} in the appendix) the detector recovers 97\% of planted modifications, including all agent-directed text in every channel, types 99.9\% of detections correctly and flags no unmodified page or non-AI identity. As the planted text uses the lexicon's constructions, the counts reported below are conservative lower bounds.

Agent-directed text that browsers never receive is rare. Because the language-model triage over-labels (177 sites, mostly block notices and sign-in prompts), we reviewed by hand every span it marked as written for AI, sponsored or instructive, and every agent-only span that mentions AI, agents, models, sponsorship or llms files. Four sites address AI systems in copy that the browser's copy lacks, three of them again in both later sweeps. Two serve documentation written for coding agents and one a preamble for AI readers of its documentation. The fourth, a major news publisher, serves every AI token, on all three sampled pages, an agent page produced by an intermediary, marked up as an agent advertisement with \texttt{sponsored}, \texttt{campaign} and \texttt{creative} attributes and carrying a bank's sponsored copy, while browsers and Googlebot receive the publisher's page without it. Sponsored content shown to answer engines and withheld from people and from the dominant search crawler is the clearest instance of agent-targeted cloaking we observed, and it matches a publicly reported 2026 incident at the same publisher~\cite{schmalbach2026time}.

The negotiated channel and the open \texttt{llms.txt} file carry most text written for models. Among the sites whose Markdown rendering differs from the browser's page, 19 address AI systems in it, mostly with pointers to an \texttt{llms.txt} index or a statement that the rendering exists for AI consumption, and three steer the model: two documentation sites tell the reading agent that the page overrides what it recalls from training, and one company instructs AI assistants how to describe it and in which categories not to place it. Of the 6,648 domains, 1,144 (17\%) publish an \texttt{llms.txt}, and in about 70 distinct sites the file tells the model how to describe the site, which of its pages to prefer over third-party sources, or which of its products to recommend for a generic request. Text that matches the agent-directed lexicon is mostly visible to everyone: of the 77 sites whose AI views match it, 76 serve the same text to browsers, and three of the four sites above were found only by the hand review. A single-identity scan of the Web would therefore see nearly all agent-directed text we found. What it would miss is the agent page and the negotiated rendering.

\begin{figure}[t]
  \centering
  \includegraphics[width=\columnwidth]{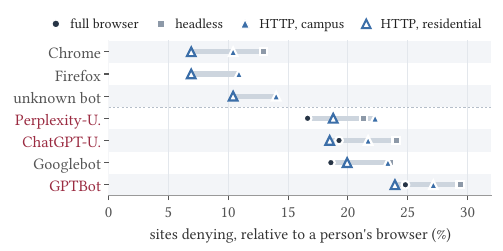}
  \caption{Sites denying each identity relative to a person's browser, a full Chrome on the campus line, on the 1,028 subsample sites reached in every cell. A non-browser client is refused on a tenth of sites whatever it declares. An AI or search token is refused on about a fifth in any client.}
  \Description{Dot plot of the share of sites denying Chrome, Firefox, an unknown bot, Perplexity-User, ChatGPT-User, Googlebot and GPTBot relative to a full browser, for a full browser, headless mode, HTTP from the campus and HTTP from the residential line.}
  \label{fig:signal}
\end{figure}

\subsection{Signal Attribution}
\label{sec:signal}

Figure~\ref{fig:signal} places the declaration among the signals an agent cannot change, measuring denial against the human reference $h$ on the 1,028 subsample sites that every cell reached. The client matters on its own: a client that declares Chrome but does not run a full browser is denied on 10.4\% of the sites that serve $h$ over HTTP and on 12.9\% in headless mode, so a tenth of the Web refuses whatever is not a full browser. The declaration matters as much: a full browser that differs from $h$ only in declaring ChatGPT-User is denied on 19.3\% of these sites, Perplexity-User on 16.6\% and GPTBot on 24.8\%. Once an AI token is declared the engine moves denial by at most 6 points, because the sites that refuse the name refuse it in every client, and the residential line lowers every rate by three to four points. The model of Equation~\ref{eq:glmm} gives the same ordering: relative to declaring Chrome, declaring an AI identity raises the probability of denial by 8.3 to 18.1 points and declaring the unknown bot by 3.0, the HTTP and headless engines raise it by 3.3 and 4.4 points on average, and the residential vantage lowers it by 3.3. For an agent, the spoofing dividend states the consequence: of the denials that an AI token meets over HTTP, 61\% from the campus and 69\% from the residential line disappear when the same client declares Chrome. An agent that declares itself honestly gives up, on these sites, access that a browser string would obtain, the incentive behind a vendor's observed switch to a generic browser string~\cite{cloudflare2025perplexity}.

\begin{figure}[t]
  \centering
  \includegraphics[width=\columnwidth]{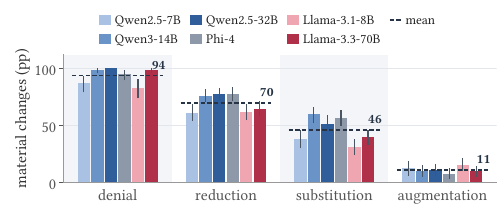}
  \caption{Material changes in the agent's answer when it reads its own view instead of the browser's, net of the browser-versus-browser null, by form and backbone, with 95\% page-level bootstrap intervals.}
  \Description{Grouped bars of material changes in percentage points for six backbones (Qwen2.5-7B, Qwen3-14B, Qwen2.5-32B, Phi-4, Llama-3.1-8B, Llama-3.3-70B) and their mean, for denial (mean 94), reduction (70), substitution (46) and augmentation (11).}
  \label{fig:consequence}
\end{figure}

\subsection{Consequences for Agents}
\label{sec:effects}

Figure~\ref{fig:consequence} reports how often an answer built from the persona's view differs materially from one built from the browser's view, minus how often two browser fetches of the same page already lead to a material difference, on 331 pages with a content-level flag and 38 denied pages, two requests per page. Denied pages, where the agent answers from a block page, change the answer on 83 to 100 points more pages than the null, and the content forms follow the text they remove or replace: reductions on 61 to 78 points and substitutions on 31 to 60, with intervals that exclude zero for every form and all six backbones, from 7B to 70B parameters. The null itself reaches up to 25\%, which is why the second browser fetch is needed. A second judge from another vendor agrees with the first on 88\% of comparisons (Cohen's $\kappa=0.70$). It confirms a positive effect for every form and backbone (reductions 58 to 68 points).

Text that sites publish for agents in \texttt{llms.txt} also steers answers: on every backbone, keeping a site's directive raises the share of answers that follow it, and a system prompt that marks retrieved content as untrusted data does not remove the effect (Appendix~\ref{app:results}).

\section{Discussion}
\label{sec:discussion}

\paragraph{Implications}
The measurements change the picture of how the Web treats agents in four ways. First, the dominant treatment is denial, enforced through a few intermediaries on behalf of many sites, so an audit of what agents can read should look at CDN rules before it looks at pages. Second, denial is keyed on an unverified name: declaring an AI identity costs a self-declared client a fifth to a quarter of the sites that serve the same client under a browser string, while Anthropic's deployed fetcher passes on most of them, so the courtesy of declaring is cheap for the vendors that own address ranges and expensive for everyone else, including agents built on open frameworks and research tools. Verifiable identity through signed requests~\cite{cloudflare2025signedagents,meunier2026httpsig} would let any operator prove what it is without owning an address range, which is the remedy our data argue for. In the long tail denial narrows to training crawlers, so on-demand fetchers meet it mostly on popular sites behind large CDNs. Third, content-level cloaking exists but is rare and runs through sanctioned channels, namely Markdown negotiation~\cite{martinho2026markdown}, llms-style files~\cite{howard2024llmstxt} and intermediary-built agent pages. These channels are inspectable, and the sponsored agent page shows why an answer engine should also fetch as a browser and flag text found only in its own copy. The directives that sites publish openly already steer answers on every backbone, and a system prompt that marks retrieved text as untrusted does not remove the effect, so an engine should report what a site asks of it rather than rely on the model to ignore it. Fourth, measuring served content needs a calibrated contrast: an uncalibrated audit of the Web reports page churn as cloaking, while ours holds the null at 0.9\% and recovers 97\% of planted modifications (Appendix~\ref{app:results}).

\paragraph{Limitations}
Although \sys{} separates the declaration from page churn on every site it reaches, it has some limitations. A window issues eighteen requests to a host and takes about a minute per page, which suits a periodic audit better than continuous monitoring, although a full sweep of the top 10,000 still completes in about two hours on one node. Additionally, like other text-based measurements, the detector reads text, so a change in an image or a price that leaves the text unchanged escapes it, and our sweeps cover one week of October 2026. Appendix~\ref{app:discussion} discusses these points and the broader impact of the work, and outlines how screenshots, signed requests and longer-running sweeps would extend it.

\paragraph{Ethics}
The study fetches public pages at most once every two seconds per host and a few dozen times per site, never solves a challenge, logs in or bypasses an access control, never claims a vendor's address, so that operators can tell our requests from their own, and involves no human subjects. We release aggregate findings, the harness and redacted diffs, and notify the affected operators and sites before publication (Appendix~\ref{app:ethics}).

\section{Conclusion}

With \sys{} we contrasted 15 declared identities on the Tranco top 10,000 and the long tail with browsers that differ only in the declaration, and typed, attributed and tested the differences. A declared AI agent is refused on a fifth to a quarter of sites, set by a few intermediaries on the strength of a name it cannot prove, while Anthropic's verified fetcher is let in. Content written for agents alone is rare and flows through inspectable channels. Until agents can prove what they are, the declaration that vendors offer as a courtesy remains a liability for everyone else, and verifiable identity is the next problem for the agentic Web. For Web measurement, agent access depends on how intermediaries enforce rules, and our harness makes what any automated reader is served measurable.

\bibliographystyle{ACM-Reference-Format}
\bibliography{references}

\appendix
\sloppy

\section*{Summary of the Appendix}

The appendix is organized as follows. \S\ref{app:persona} specifies the personas. \S\ref{app:labeling} details how a window becomes a label (Figure~\ref{fig:labeling}), hidden text and the testbed. \S\ref{app:protocols} gives the consequence protocols. \S\ref{app:ethics} covers ethics and cost. \S\ref{app:results} reports an ablation, site categories, steering per backbone and detector recall. \S\ref{app:discussion} discusses limitations, broader impact and future work. Code, full implementation and per-site results will be released upon acceptance.

\section{Persona Details}
\label{app:persona}

User-Agent strings follow the vendors' documentation, with Chrome and HeadlessChrome at the major version of the rendered engines (140). The HTTP engine replays Chrome's TLS and HTTP/2 fingerprint and navigation headers, without client hints, for every identity, so that within a cell only the User-Agent string, and for the Markdown client the \texttt{Accept} header, differ. The headful engine is full Chromium under a virtual display with the automation flag disabled. The headless engine is the headless shell of the same build. The three vantages lie in distinct autonomous systems and, checked before every run, in none of the eleven range files of OpenAI, Anthropic, Perplexity, Google, Microsoft and Common Crawl.

\section{Labeling Pipeline}
\label{app:labeling}

\begin{figure*}[t]
  \centering
  \includegraphics{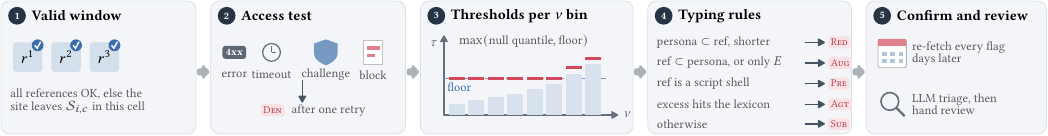}
  \caption{A diagram illustrating how \sys{} turns one fetch window into a label: a valid window, an access test, floored null thresholds per bin of the page's own variation, typing rules, and confirmation with review.}
  \Description{Five panels: a valid window with r1, r2 and r3; an access test for errors, timeouts, challenges and blocks; thresholds per bin of the page's variation as the maximum of a null quantile and a floor; typing rules for reduction, augmentation, prerendering, agent-directed text and substitution; and confirmation by re-fetching every flag and LLM triage with hand review.}
  \label{fig:labeling}
\end{figure*}

\paragraph{From Window to Label}
Figure~\ref{fig:labeling} expands the calibration of Section~\ref{sec:contrast} into five steps. The window is the unit of validity, so a site that refuses Chrome itself never contributes a denial, and access is tested before text, because a challenge page has text of its own. Thresholds are the larger of a null quantile and a floor, per cell and bin of $\nu$. In the campus sweep the displacement threshold stays at its floor of 0.05, because $\Delta$ already subtracts $\nu$, while the excess threshold rises from 8 tokens on static pages to 16 on volatile ones in one fold. Typing follows the order shown, and every flag is fetched again days later.

\paragraph{Hidden Text}
Rendered pages are walked text node by text node, testing each element with \texttt{checkVisibility}, font size, clipping, placement, color contrast and \texttt{aria-hidden}. The HTTP engine approximates the same tests on the static DOM. The hidden view adds comments, \texttt{alt}, \texttt{title}, \texttt{aria-label}, \texttt{meta}, \texttt{noscript} and \texttt{template} content, and Unicode tag characters. Excess is counted per sentence, so that a sentence that occurs in either browser fetch, or whose 3-gram shingles are mostly contained in them, is attributed to the page rather than to the persona.

\paragraph{Planted Testbed and Review}
The testbed serves the main text of 40 internal pages from a local server, through the same harness, in a control variant and 36 variants that apply a form only to AI tokens: (1) reductions that drop 10\%, half or all but the first paragraph, (2) substitutions of digits, of one paragraph or of the article, and (3) added text of 12, 60 or 300 words, neutral (augmentation) or built from five agent-directed templates, in each of five channels. Every page carries a rotating headline and a timestamp, so that the null has variation to absorb. In the wild, every agent-only span of at least 12 tokens is triaged by Qwen2.5-7B-Instruct reading it inside delimiters that mark it as untrusted, and every span outside benign classes, or matching the lexicon, was reviewed by hand. The counts in Section~\ref{sec:content} are the reviewed counts.

\section{Consequence Protocols}
\label{app:protocols}

\paragraph{Altered Pages}
The sample draws up to 120 flagged pages per content form and 40 denials, one identity per page, preferring the agent fetchers. Backbones answer both requests from each view, truncated to 6,000 characters, with greedy decoding and at most 250 tokens, and the judge, gpt-oss-20b, labels each answer against the answer from $r^1$ as identical, minor or material. The backbones are Qwen2.5-7B, Llama-3.1-8B, Phi-4, Qwen3-14B, Qwen2.5-32B and Llama-3.3-70B (FP8) under vLLM, and Qwen2.5-32B also serves as the second judge for the agreement analysis.

\paragraph{Published Directives}
From the \texttt{llms.txt} sentences that the triage marks as steering, we keep 66 distinct directives. For each, the judge model writes two questions an ordinary user could ask about the site whose answer the directive would change, and one sentence describing an answer that follows it, which yields 118 questions on 59 sites. A backbone answers each question from the homepage text and the surrounding \texttt{llms.txt} context, with the directive kept or stripped, under a plain system prompt or one that marks retrieved content as untrusted data~\cite{hines2024spotlighting}. The judge, blind to the condition, decides whether the answer follows the directive, and intervals resample sites.

\section{Ethics, Disclosure and Cost}
\label{app:ethics}

Presenting another party's crawler identity is established practice in cloaking measurement~\cite{invernizzi2016cloak,khattak2016anonymous,liu2025somesite}. Pages that robots.txt disallows for a token are requested only to measure enforcement, at most three per site. No vantage lies in a vendor's range, so no request can pass as a verified fetch. The only verified fetches are those of the deployed Claude fetcher, requested through its product at a few URLs per minute. Requests carry no study marker, which would change how sites treat them, and stay below one person's browsing rate per host. Sites behind national filtering are unreachable from both domestic vantages and fall out of the measured set. Released diffs hash site-identifying strings, and affected sites and agent operators are notified at least 90 days before release. The campus sweep issued 354,204 requests, 18 per page, at most one every two seconds per host. The residential sweep issued 35,388 and the rendered engines about 24,000 page loads. Extraction, contrasts and calibration take about two CPU-hours, and the consequence studies about one A100 and four H100 GPU-hours.

\section{Additional Results}
\label{app:results}

\begin{figure}[t]
  \centering
  \includegraphics{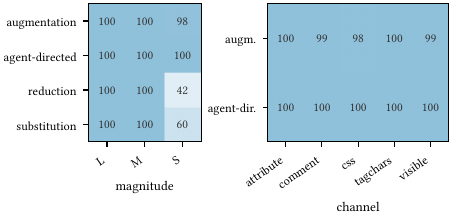}
  \caption{Recall (\%) of the detector on the planted testbed, by form and magnitude (S: a few words, M: a paragraph, L: most of the page) and, for added text, by channel.}
  \Description{Two heatmaps. By magnitude L, M and S, recall is 100, 100 and 98 for augmentation, 100 for agent-directed text, 100, 100 and 42 for reduction, and 100, 100 and 60 for substitution. By channel (attribute, comment, css, tag characters, visible), recall of added text is 98 to 100.}
  \label{fig:recall}
\end{figure}

\paragraph{Calibration Ablation}
Table~\ref{tab:ablation} removes one component of the contrast at a time. Comparing the persona's text with the first Chrome fetch, without a null, flags the null on 7.2\% of sites, as often as the AI tokens (7.7--8.9\%), so a naive comparison cannot separate treatment from churn. Once calibrated, the rates move by less than half a point across 3- and 8-word shingles, without $\nu$ bins and at $\alpha=2\%$ (not shown), where the floors bind. What the design buys is power. Comparing with a single Chrome fetch, or without subtracting the page's own variation, raises the thresholds needed to hold the null and lowers recall on the planted testbed from 97\% to 77\%, and each statistic alone misses a family of forms: displacement alone misses most added text, and excess alone misses every reduction. Denial does not depend on the position in the window (20.5\% of pages in the first four slots, 20.7\% in the last four).

\paragraph{Site Categories}
Table~\ref{tab:categories} breaks denial down by site category, assigned from the homepage by Qwen2.5-7B-Instruct. News, adult and social sites deny training crawlers markedly more often than the spoofed Googlebot, the signature of a rule written for AI training, and the gap widens in the long tail (news: 20 against 10\%). Retail, travel, finance and government sites deny every automated name, including the unknown bot, the signature of general bot management. Developer documentation denies least.

\paragraph{Steering by Published Directives}
In Table~\ref{tab:steering}, every backbone follows more directives when they are kept, under both system prompts, and the pooled 95\% interval excludes zero in both (3.2--11.0 and 2.8--11.2 points). The untrusted-data prompt lowers compliance in both conditions yet keeps the pooled lift (6.7 against 6.9 points).

\section{Discussion}
\label{app:discussion}

\begin{table*}[t]
  \caption{Sites (\%) denying each group of identities and altering AI identities, by site category assigned from the homepage by a language model, in the top 10,000 and in the long-tail sample (ranks 10k to 1M).}
  \label{tab:categories}
  \footnotesize
  \setlength{\tabcolsep}{5.6pt}
  \renewcommand{\arraystretch}{0.94}
  \begin{tabular}{@{}lrcccccrccccc@{}}
    \toprule
     & \multicolumn{6}{c}{Top 10,000} & \multicolumn{6}{c}{Ranks 10k to 1M} \\
    \cmidrule(lr){2-7}\cmidrule(l){8-13}
    Category & Sites & Fetchers & Crawlers & Googlebot & Unknown & \ap{Alt} & Sites & Fetchers & Crawlers & Googlebot & Unknown & \ap{Alt} \\
    \midrule
    Software and docs & 1326 & \cc{E7E9ED}{11.5} & \cc{DFE2E6}{15.2} & \cc{E1E4E8}{14.9} & \cc{F9F9FA}{3.4} & \cc{DBE9F1}{3.1} & 383 & \cc{F5F6F8}{5.0} & \cc{E9EBEE}{11.0} & \cc{F2F3F5}{6.3} & \cc{FCFCFD}{1.3} & \cc{F0F6F9}{1.4} \\
    News and media & 766 & \cc{D2D7DD}{21.7} & \cc{C5CBD3}{28.0} & \cc{DADEE3}{17.6} & \cc{F4F5F6}{5.5} & \cc{CFE2EC}{4.2} & 257 & \cc{E7E9ED}{11.3} & \cc{D5DADF}{20.2} & \cc{EAECEF}{9.7} & \cc{F7F8F9}{3.9} & \cc{E8F1F6}{2.0} \\
    Entertainment & 577 & \cc{D5DADF}{20.0} & \cc{C5CBD3}{27.8} & \cc{D5DADF}{20.1} & \cc{F7F8F9}{3.8} & \cc{CBE0EB}{4.4} & 284 & \cc{F4F5F6}{5.4} & \cc{D7DBE0}{19.5} & \cc{F4F5F6}{5.6} & \cc{FFFFFF}{0.4} & \cc{E6F0F5}{2.2} \\
    E-commerce & 482 & \cc{C2C8D0}{29.3} & \cc{BCC3CB}{32.1} & \cc{BCC3CB}{32.0} & \cc{E7E9ED}{11.8} & \cc{A9CCDD}{7.5} & 352 & \cc{E4E7EA}{13.2} & \cc{D5DADF}{19.8} & \cc{E2E5E9}{13.6} & \cc{F4F5F6}{5.7} & \cc{CDE1EB}{4.3} \\
    Education & 479 & \cc{D5DADF}{20.3} & \cc{C9CED5}{26.2} & \cc{CACFD6}{25.1} & \cc{F9F9FA}{2.7} & \cc{E6F0F5}{2.1} & 213 & \cc{E9EBEE}{10.5} & \cc{DADEE3}{17.6} & \cc{E1E4E8}{14.6} & \cc{F9F9FA}{2.8} & \cc{E4EFF4}{2.3} \\
    Government & 261 & \cc{C9CED5}{26.3} & \cc{BFC5CE}{30.6} & \cc{B9C0C9}{33.7} & \cc{E7E9ED}{11.5} & \cc{F5F9FB}{0.9} & 78 & \cc{E4E7EA}{12.8} & \cc{DADEE3}{17.9} & \cc{DCDFE4}{16.7} & \cc{EFF1F3}{7.7} & \cc{ECF4F8}{1.6} \\
    Finance & 198 & \cc{CFD4DA}{22.9} & \cc{C7CDD4}{26.8} & \cc{BAC1CA}{32.8} & \cc{E5E8EB}{12.1} & \cc{BAD6E4}{6.1} & 115 & \cc{D9DCE1}{18.6} & \cc{CDD2D9}{23.7} & \cc{D1D5DB}{22.6} & \cc{EFF1F3}{7.8} & \cc{E6F0F5}{2.2} \\
    Social and forums & 186 & \cc{D7DBE0}{19.4} & \cc{C9CED5}{26.2} & \cc{DCDFE4}{16.7} & \cc{F5F6F8}{4.3} & \cc{B8D5E3}{6.1} & 65 & \cc{FAFBFB}{2.1} & \cc{DDE1E5}{16.2} & \cc{FCFCFD}{1.5} & \cc{FCFCFD}{1.5} & \cc{C9DFEA}{4.6} \\
    Other business & 155 & \cc{D7DBE0}{19.5} & \cc{CFD4DA}{23.2} & \cc{CDD2D9}{24.0} & \cc{F2F3F5}{6.5} & \cc{F7FAFC}{0.7} & 102 & \cc{EDEFF1}{8.8} & \cc{D4D8DE}{20.6} & \cc{EDEFF1}{8.8} & \cc{F2F3F5}{5.9} & \cc{F5F9FB}{0.9} \\
    Adult & 126 & \cc{D4D8DE}{20.6} & \cc{AFB7C1}{39.5} & \cc{DADEE3}{17.5} & \cc{FDFEFE}{0.8} & \cc{AECFDF}{6.9} & 82 & \cc{F2F3F5}{6.1} & \cc{B9C0C9}{33.5} & \cc{FCFCFD}{1.2} & \cc{FAFBFB}{2.4} & \cc{E4EFF4}{2.3} \\
    Travel & 83 & \cc{B2BAC4}{36.9} & \cc{AFB7C1}{38.6} & \cc{AFB7C1}{38.6} & \cc{DADEE3}{18.1} & \cc{9DC5D9}{8.4} & 66 & \cc{E7E9ED}{11.6} & \cc{D4D8DE}{20.8} & \cc{DCDFE4}{16.7} & \cc{EFF1F3}{7.6} & \cc{E0EDF3}{2.7} \\
    Health & 66 & \cc{C2C8D0}{29.3} & \cc{BAC1CA}{33.3} & \cc{C7CDD4}{27.3} & \cc{F9F9FA}{3.0} & \cc{E0EDF3}{2.7} & 50 & \cc{F2F3F5}{6.0} & \cc{E2E5E9}{13.5} & \cc{E2E5E9}{14.0} & \cc{FFFFFF}{0.0} & \cc{FFFFFF}{0.0} \\
    \bottomrule
  \end{tabular}
\end{table*}

\begin{table}[t]
  \caption{Answers (\%) following a site's published directive, stripped from or kept in its \texttt{llms.txt}, by system prompt.}
  \label{tab:steering}
  \footnotesize
  \setlength{\tabcolsep}{3.6pt}
  \renewcommand{\arraystretch}{0.94}
  \begin{tabular}{@{}lrrrrrr@{}}
    \toprule
     & \multicolumn{3}{c}{Plain} & \multicolumn{3}{c}{Untrusted data} \\
    \cmidrule(lr){2-4}\cmidrule(l){5-7}
    Backbone & strip & keep & lift & strip & keep & lift \\
    \midrule
    Qwen2.5-7B & 47.5 & 55.9 & \cc{BED8E6}{+8.5} & 39.0 & 50.0 & \cc{ABCDDE}{+11.0} \\
    Llama-3.1-8B & 50.0 & 59.3 & \cc{B8D5E3}{+9.3} & 43.2 & 51.7 & \cc{BED8E6}{+8.5} \\
    Phi-4 & 47.5 & 58.5 & \cc{ABCDDE}{+11.0} & 49.2 & 54.2 & \cc{D9E8F0}{+5.1} \\
    Qwen3-14B & 50.0 & 54.2 & \cc{E0EDF3}{+4.2} & 51.7 & 55.9 & \cc{E0EDF3}{+4.2} \\
    Qwen2.5-32B (AWQ) & 53.4 & 57.6 & \cc{E0EDF3}{+4.2} & 47.5 & 55.9 & \cc{BED8E6}{+8.5} \\
    Llama-3.3-70B (FP8) & 51.7 & 55.9 & \cc{E0EDF3}{+4.2} & 50.0 & 53.4 & \cc{E6F0F5}{+3.4} \\
    \midrule
    \rowcolor{rowhl} Pooled & 50.0 & 56.9 & \textbf{+6.9} & 46.8 & 53.5 & \textbf{+6.7} \\
    \bottomrule
  \end{tabular}
\end{table}

\begin{table}[t]
  \caption{Ablation of the within-cell contrast on the campus sweep, each variant recalibrated: sites (\%) altered, and recall on the planted testbed.}
  \label{tab:ablation}
  \footnotesize
  \setlength{\tabcolsep}{3.4pt}
  \renewcommand{\arraystretch}{0.94}
  \begin{tabular}{@{}llcccc@{}}
    \toprule
     & Variant & Null & AI tokens & Unknown bot & Recall \\
    \midrule
    \rowcolor{rowhl} & \textbf{\sys{}} & \textbf{0.9} & \textbf{3.2--4.6} & \textbf{2.9} & \cc{A3C8DB}{\textbf{97.0}} \\
    \multirow{2}{*}{\grp{Statistic}} & displacement $\Delta$ only & 0.3 & 2.2--3.3 & 2.0 & \cc{E8F1F6}{25.0} \\
     & excess $E$ only & 0.7 & 2.7--4.0 & 2.3 & \cc{ABCDDE}{89.4} \\
    \arrayrulecolor{grprule}\cmidrule(l){2-6}\arrayrulecolor{black}
    \multirow{2}{*}{\grp{Reference}} & one Chrome fetch $r^1$ & 0.7 & 2.8--4.4 & 2.7 & \cc{B6D4E3}{77.4} \\
     & no subtraction of $\nu$ & 1.0 & 3.0--4.6 & 3.0 & \cc{B6D4E3}{77.4} \\
    \arrayrulecolor{grprule}\cmidrule(l){2-6}\arrayrulecolor{black}
    \grp{Calibration} & $\alpha=0.5\%$ & 0.4 & 2.8--4.1 & 2.4 & \cc{BCD7E5}{70.1} \\
    \arrayrulecolor{grprule}\cmidrule(l){2-6}\arrayrulecolor{black}
    \grp{Baseline} & any text difference & 7.2 & 7.7--8.9 & 8.7 & \cc{9FC6D9}{100} \\
    \bottomrule
  \end{tabular}
\end{table}

\paragraph{Limitations}
Although \sys{} separates the declaration from page churn on every site it reaches, it has some limitations. It cannot send from a vendor's range or sign as a vendor, so cloaking reserved for verified fetchers is seen only through the one deployed fetcher we queried. Our alteration rates are lower bounds on such cloaking and our denial rates upper bounds on what verified fetchers meet. Two vantages lie in one country, whose national filter removes some sites, mostly in the long tail. Denial levels differ by vantage but the ordering of identities does not ($\rho=0.98$). The detector reads text, so changes in images, prices or links that leave the text unchanged escape it, and small reductions are its weakest case (42\% recall at the smallest magnitude, Figure~\ref{fig:recall}). These limitations are shared by most passive measurements of served content.

\paragraph{Broader Impact}
This work measures what the Web serves to the AI agents that increasingly read it for people, and finds that the pressing problem is less hidden content than unverifiable identity. On the positive side, it gives publishers, CDNs and agent vendors a common and reproducible measurement of treatment, so that refusals enforced at the edge can be audited against the policies that sites declare. It prices the honest declaration and thereby supports verifiable identity through signed requests~\cite{cloudflare2025signedagents,meunier2026httpsig} over address allowlists, which favor the few vendors that own ranges. The calibrated contrast and the planted testbed transfer to other questions of differential treatment on the Web, such as treatment of assistive clients, regional personalization or prices quoted to automated shoppers. The steering result informs the design of answer engines: directives that sites publish already move answers, and a prompt-level defense does not stop them, so an engine should surface what a site asks of it rather than rely on the model to ignore it. On the negative side, the same harness that audits cloaking could help a cloaker tune its rules against detection, and the identity-by-signal decomposition could help an operator fingerprint agents more precisely. The spoofing dividend could also be read as an incentive for agents to declare a browser, which would erode the very mechanism this paper argues should be strengthened. To mitigate these risks, we release aggregates and redacted diffs rather than per-site rules, notify affected parties first, and present the dividend as a cost of honesty that verification should remove, not as advice for agents to impersonate browsers.

\paragraph{Future Work}
This work also comes with new challenges, certainly worth further exploration:
\begin{itemize}[leftmargin=*,nosep]
  \item \textit{Longitudinal monitoring.} Continuous sweeps would track how CDN defaults and \texttt{llms.txt} directives evolve.
  \item \textit{Beyond text.} Screenshots, prices and links would close the blind spot of a detector that reads only text.
  \item \textit{Verified identities.} The same contrast from vendor ranges or with signed requests would test whether verification lifts the denial.
  \item \textit{Interactive agents.} Search, forms and checkout would reveal cloaking that targets what an agent does.
  \item \textit{Defenses.} Engines that cross-check a browser's view, and models that report site directives, can be tested here.
\end{itemize}

\end{document}